\documentclass[prl,showpacs,preprintnumbers,amsmath,amssymb,endfloats]{revtex4}

\usepackage{graphicx}
\usepackage{dcolumn}
\usepackage{bm}

\begin{document}


\title{Non--Ginsburg-Landau Critical Current Behavior in MgCNi$_3$}

\author{D.P. Young, M. Moldovan, and P.W. Adams}
\affiliation{Department of Physics and Astronomy\\Louisiana State University\\Baton Rouge, Louisiana,
70803}%


\date{\today}

\begin{abstract}
We present transport critical current measurements on microfibers consisting of a 80-nm thick layer of polycrystalline MgCNi$_3$
synthesized directly onto 7-$\mu$m diameter carbon fibers.  Near the transition temperature, $T_c$, the critical current density, $J_c$, is
well described by the power law form $\left[1-(T/T_c)^2)\right]^\alpha$, where $\alpha=2$ with no crossover to the Ginsburg-Landau exponent
$\alpha=1.5$.   We extrapolate $J_c(0)\approx4$ x $10^7$ A/cm$^2$, which is an order of magnitude greater than estimates obtained from
magnetization measurements of polycrystalline powders.  The field dependence is purely exponential $J_c(T,H)=J_c(T)\exp(-H/H_o)$ over the
entire field range of 0 to 9 T.  The unconventional scaling behavior of the critical current appears rooted in an
anomalous temperature dependence of the London penetration depth, suggesting an unusual superconducting ground state in MgCNi$_3$.
           
\end{abstract}

\maketitle

The only known superconducting non-oxide perovskite MgCNi$_3$ has come under intense study since its discovery two years ago \cite{MgCNi3}. 
It is widely believed to be an important analog to the high-$T_c$ perovskites by virtue of its chemical composition and its non-layered
cubic structure.  One would hope that MgCNi$_3$ could help disentangle the influences of crystal symmetry, chemical
doping, and micro-morphology in its oxide cousins.  Furthermore, band structure calculations have indicated that the
superconducting phase of MgCNi$_3$ may, in fact, be near a ferromagnetic ground state, which is in accord with its high Ni content
\cite{MgCNi3BS}.  If this is indeed the case, then it would be related to the newly discovered non-conventional itinerate ferromagnetic
superconductors UGe$_2$ \cite{UGe2} and ZrZn$_2$ \cite{ZrZn2}.  Unfortunately, the volatility of Mg has hampered the synthesis of bulk
crystalline samples of MgCNi$_3$, and, up to quite recently, only polycrystalline powders have been available \cite{MgCNi3films}.  
Consequently, measurements of the transport and electron tunneling properties of the material have been compromised by poor sample
morphology, making a quantitative characterization of the superconducting phase difficult and, in particular, a determination of the pairing
symmetry uncertain at best.  In the present Letter we report the first transport critical current measurements of
annular MgCNi$_3$ fibers as a function of temperature and magnetic field.  We show that that the scaling behavior near the transition
temperature,
$T_c$, is well described by a non--Ginsburg-Landau quadratic form \cite{Tinkham} and that the overall magnitude of the zero-field critical
current density, $J_c(0)$, is in good agreement with that determined from estimates of the thermodynamic critical field, $H_c$, and the
London penetration depth, $\lambda$ \cite{MgCNi3data}.  The critical current is exponentially suppressed by axial magnetic field over the
entire field range 0-9 T.

 The annular MgCNi$_3$ fibers were formed by reacting Ni-coated carbon fibers in excess Mg vapor at 700 $^o$C for 20-30 minutes in a
vacuum sealed quartz tube.  This process is similar to that used to form MgCNi$_3$ films, where CNi$_3$ precursor films were exposed to Mg
vapor
\cite{MgCNi3films}. The Ni-coated carbon fibers used in this study were obtained from Novamet Specialty Products Corporation under the
product name Incofiber $12K20$ \cite{Novamet}.  Incofiber $12K20$ consists of 6-8 micron diameter carbon fibers which are coated with a
80-nm thick film of Ni (99.97\%) via a chemical vapor deposition process based on the high temperature decomposition of Ni(CO)$_4$.  This
deposition method is ideal for the critical current studies described below in that it tends to produce quite uniform coatings and excellent
adhesion.  We also studied fibers in which the Ni coating was produced by an electrochemical deposition process and obtained similar
results.  Annular contacts were made to the reacted fibers using Epotek conductive epoxy.  Critical currents were measured in a 4-probe
geometry using a standard pulsed technique.  Currents were driven using pulse durations of 1-2 $\mu$s with a duty cycle of 1/1000, and the
resulting voltages were measured via a boxcar integrator.  Care was taken to ensure that the pulse width and duty cycle were low enough to
avoid significant Joule heating at the contacts.  The samples were cooled by vapor down to 1.8 K in magnetic fields up to 9 T via a Quantum
Design PPMS. 

	Shown in Fig. 1 is the low current density temperature dependence of the resistivity of a 7-$\mu$m diameter pristine Ni-coated fiber and a
reacted fiber (see Fig.\ 1 inset).   The resistivities have been normalized by their room temperature values.  Note the residual resistivity
ratio of the Ni and the MgCNi$_3$ fibers are comparable and that the normal state temperature dependence of the MgCNi$_3$ fibers appears to
be logarithmic.  The midpoint resistive transition temperature of the 80 nm-thick MgCNi$_3$ layer is $T_c=7.8$ K.  This $T_c$ is slightly
higher than that of bulk powders and that recently reported in 60 nm-thick polycrystalline films on sapphire
\cite{MgCNi3films}.  The transition widths in the fibers are approximately 1/3 that of comparably thick planar films and the extrapolated
upper critical field $H_{c2}(0)\sim16.5$ T is somewhat higher than that of films ($H_{c2}(0)\sim13$ T).  Critical current measurements were
limited to temperatures above 5 K due to both the limitation of the electronics and the risk of damaging the samples.  In Fig.\ 2 we present
a log-log plot of the critical current density in zero magnetic field as a function of reduced temperature, where we have defined $J_c$ by
the onset of voltage.  The value of $T_c$ used to scale the data in Fig.\ 2 was determined by the onset of voltage at
$J=0.01J_c$.  The dashed line in Fig.\ 2 is the Ginsburg-Landau (G-L) critical current behavior for a homogeneous order parameter
\cite{Tinkham},

\begin{equation} 
J_c=\frac{H_c(T)}{3\sqrt{6}\pi\lambda(T)}\propto\left[1-(T/T_c)^2\right]^{3/2},
\end{equation} 
where $H_c$ is the thermodynamic critical field and $\lambda$ is the London penetration depth.  Early critical current measurements in
elemental films such as Sn and Pb showed good agreement with Eq.(1),  with deviations attributed to edge effects \cite{GLCC}.  Typically,
the magnitude of the critical current density in elemental films is $J_c(0)>10^7$ A/cm$^2$.  In order to rule out possible systematic errors
arising from our technique, we measured the critical current behavior of a 200-nm thick Pb film deposited via e-beam evaporation onto a
7-$\mu$m diameter rotating carbon fiber.  The temperature dependence and the overall magnitude of the critical current of the Pb-coated
fibers were found to be in good agreement with Eq.(1).  

	Significant deviations from the 3/2 exponent of Eq.(1) have been reported in measurements of
high-$T_c$ oxides, where it is not unusual for the extrapolated values of
$J_c(0)$ to be an order of magnitude lower than those of metal films.  In particular, non--Ginsburg-Landau scaling behavior has been
reported in La-Sr-Cu-O films
\cite{LaSrCuO}, Bi-Ca-Sr-Cu-O films \cite{BCSCO}, oriented YBa$_2$Cu$_3$O$_{7-y}$ films \cite{YBCO}, and YBa$_2$Cu$_3$O$_7$ bicrystals
\cite{YBCObc}.  The discrepancies have been mostly attributed to weak links that are thought to be intrinsic to the complex granular
microstructure of these ceramic materials.  Observed temperature dependencies of the critical current in the high-$T_c$ oxides are usually
of the form $J_c(T)=J_c(0)(1-T/T_c)^\alpha$ near $T_c$, with $\alpha$ ranging between 1 and 2.  (Note that near $T_c$,
$(1-(T/T_c)^2)\sim2(1-T/T_c)$).  Models of critical current behavior in granular systems assume that the intergrain transport is dominated
by tunneling processes associated with either superconductor-insulator-superconductor (SIS) Josephson junctions \cite{AB} or
superconductor-normal-superconductor (SNS) proximity junctions\cite{deGennes}.   The former is expected to produce a $(1-T/T_c)$ scaling and
the latter $(1-T/T_c)^2$.  Interestingly, granular systems can exhibit crossover behavior from one power law exponent to another
\cite{crossover}.  For instance, evidence for SIS tunneling in thin YBCO films is seen in a critical current temperature dependence that
evolves from the $\alpha=1$ of Josephson coupling to the Ginsburg-Landau $\alpha=3/2$ as one approaches $T_c$ \cite{YBCOcross}.  A similar
crossover behavior can be induced by the application of a modest magnetic field ($\sim0.01$ T), where the G-L scaling is recovered in finite
field \cite{YBCOcross}.  

	The data in Fig.\ 2 show no evidence of crossover behavior, suggesting that
the quadratic dependence is intrinsic and not an artifact of the micromorphology of the fibers.  Furthermore, using the reported estimates of
the zero-temperature thermodynamic field and penetration depth \cite{MgCNi3data}, $\mu_oH_c(0)=0.22$ T and $\lambda(0)=220$ nm in Eq.(1), one
obtains a reasonably large value of $J_c(0)\approx4$ x $10^7$ A/cm$^2$ which is in good agreement with the zero-temperature extrapolation
of the data in Fig.\ 2.  This suggests that the critical current magnitude is not compromised by weak links.   

	In Fig.\ 3 we have plotted the onset transition temperature shift as a function of current density.  These data are presented as a
consistency check of the $\alpha=2$ power law behavior in Fig.\ 2.   The solid line is a guide to the eye and has the expected slope of
1/2.  The dashed line represents the quadratic Joule heating dependence of which there is no indication in the data.  If the anomalous
scaling of figures 2 and 3 is, in fact, a property of the condensate, then it seems likely that it originates in the temperature dependence
of the penetration depth in Eq.(1).  In terms of G-L theory \cite{Tinkham},

\begin{equation}
\lambda^2=\frac{m}{2\mu_o|\psi|^2e^2}
\end{equation}
where $m$ is the electron mass, $e$ the electron charge, $\mu_o$ the vacuum permeability, and $\psi$ the condensate wavefunction, whose
square modulus is proportional to the density of Cooper pairs, $n_s$.  The theory assumes that near $T_c$, $|\psi|^2\propto(1-T/T_c)$. 
However, if the wavefunction couples to fluctuations of some other order parameter, such as that of a ferromagnetic phase, then critical
behavior can be altered.  In the case of MgCNi$_3$, it is believed that van Hove singularities in the Ni 3d bands may lead to paramagnon
fluctuations that could influence the nature of the pairing \cite{MgCNi3BS}.  The quadratic scaling behavior of the data in Fig.\ 2 is,
in fact, consistent with a linear power law $\lambda\propto (1-(T/T_c)^2)$, in contrast to the square-root law of Eq.(2),
perhaps supporting such a scenario.  The case for an anomalous penetration depth in MgCNi$_3$ has recently been strengthened by
low temperature measurements \cite{Prozorov} in which a distinctly non-BCS quadratic temperature
dependence is reported, $\lambda(T)\propto T^2$ for $T<T_c/4$, consistent with our analysis. The $T^2$ dependence can be interpreted
as evidence for a nodal gap structure such as that associated with $d$-wave superconductivity. 
However, the issue of the pairing state in MgCNi$_3$ remains controversial in that other probes such as electron tunneling and heat capacity
are consistent with conventional BCS phonon-mediated superconductivity
\cite{MgCNi3data,BCSpairing}.  Nevertheless, there is still an open question as to whether or not a nearby ferromagnetic ground state
exists, and, if so, to what extent it influences the superconducting properties of the system. 

	In Fig.\ 4 we plot the critical current density as a function of magnetic field at several different temperatures.  The magnetic
field was applied axially to the fibers.  Note that the exponential field dependence exists over the entire field range.  The
solid lines are exponential fits to the data from which a characteristic field $H_o$ is extracted, see Fig.\ 4 inset.  Interestingly,
virtually no hysteresis in either field or current was observed during the course of these measurements, suggesting that flux pinning was
not significant. In contrast, strong pinning effects have been reported in magnetization measurements of polycrystalline powders of
MgCNi$_3$ \cite{MgCNi3pinning}, which have been attributed to superconducting vortex core pinning
on intragrain graphite nanoprecipitates.  It seems unlikely that the MgCNi$_3$ sheath on our fibers contains unreacted
islands of carbon.  We believe that since the reaction occurs in excess Mg vapor, carbon is
leached out of the fiber to form MgCNi$_3$ at the correct stoichiometry until the nickel is completely consumed.  The higher $T_c$'s, sharp
transitions, and higher critical fields of the fibers seem to support this scenario, providing further evidence that the anomalous critical
current behavior is an intrinsic property.

	Though exponential suppression of $J_c$ has also been reported in a number of high-T$_c$ systems, they generally only show an exponential
dependence over a limited field range.  The $+$ symbols in Fig.\ 4 are critical current values obtained from magnetization measurements of
MgCNi$_3$ powders \cite{MgCNi3pinning}.  Though they are approximately an order of magnitude smaller than our transport values, they
nevertheless display a very similar exponential dependence.  In the inset of Fig.\ 4 we plot the inverse of the characteristic field as a
function of the inverse of the reduced temperature.  The dashed line in the inset of Fig.\ 4 is a linear fit to the data which
gives a slope whose value is approximately twice that of the intercept.  The fit implies the following empirical expression for the
temperature dependence of the characteristic field,
\begin{equation}
H_o(T)\approx2.5\left[\frac{1-T/T_c}{3/2-T/T_c}\right]
\end{equation} where $H_o$ is in units of Tesla.  The linear scaling behavior of Eq.(3) may be related to that of the critical current. 
Interestingly, the asymptotic behavior of $H_o$ near $T_c$ is the same as that of the thermodynamic critical field $H_c$.  However, the
magnitude of $H_o$ is much larger than $H_c$.

	In conclusion, the temperature and field dependence of the transport critical current in MgCNi$_3$ microfibers is well
described by $J_c(T,H)=J_c(0)\left[1-(T/T_c)^2\right]^2\exp[-(H/H_o)]$, where $J_c(0)\approx4$ x $10^7$ A/cm$^2$.  We believe that the
non--Ginsburg-Landau exponent is not a morphological effect and that it may be reflective of the anomalous penetration depth behavior
recently reported in polycrystalline powders \cite{Prozorov}.  Whether or not the anomalous temperature dependence of the London
penetration depth can be attributed to a non-conventional pairing mechanism in MgCNi$_3$ remains controversial.  But even
if the superconducting state is conventional, the critical behavior of MgCNi$_3$, as manifest through the critical current
scaling, is not.     

We gratefully acknowledge discussions with Dana Browne, Ilya Vekhter, Dietrich Belitz, Milind Kunchur, and Lisa Podlaha. This work was
supported by the National Science Foundation under Grant DMR 02-04871 and the LEQSF under Grant 2001-04-RD-A-11.

\newpage
\begin{figure}
\includegraphics[width=5.in]{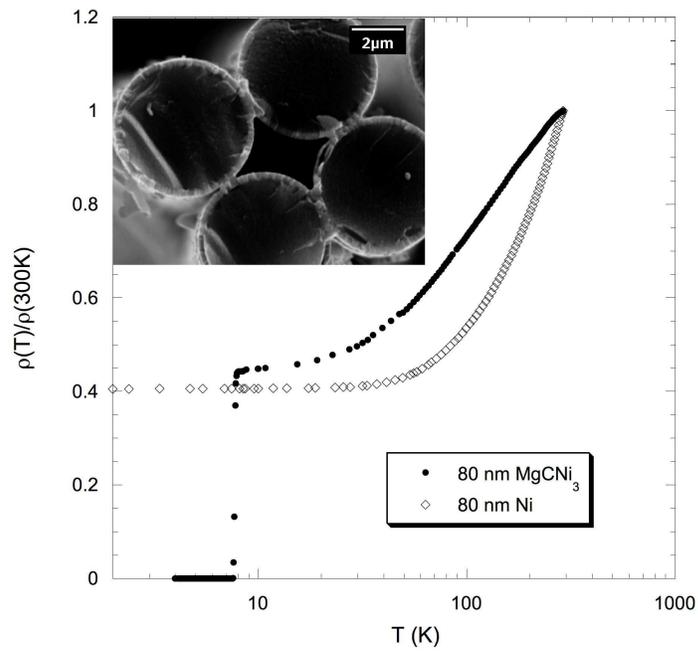}
\caption{\label{fig:epsart} Resistivity normalized by its room temperature value as a function of temperature for a carbon fiber coated with
80-nm of Ni and a similar Ni-coated carbon fiber processed in Mg vapor to form a 80-nm MgCNi$_3$ sheath.  Inset:  Scanning electron
micrograph of MgCNi$_3$ coated carbon fibers showing a cross-sectional view.}
\newpage
\end{figure}

\begin{figure}
\includegraphics[width=5.in]{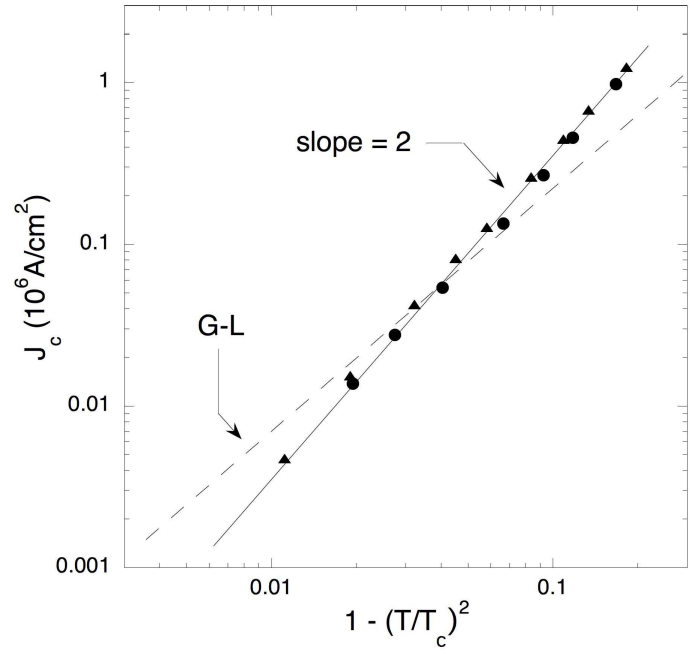}
\caption{\label{fig:epsart}  Log-log plot of the scaling behavior of the critical current density near T$_c$ in zero magnetic field for two
80-nm fibers.  The solid line is provided as a guide to the eye and extropolates to a zero-temperature density J$_c(0)=4$ x $10^7$
A/cm$^2$.  The dashed line has the Ginsburg-Landau slope of 3/2.}
\end{figure}

\begin{figure}
\includegraphics[width=5in]{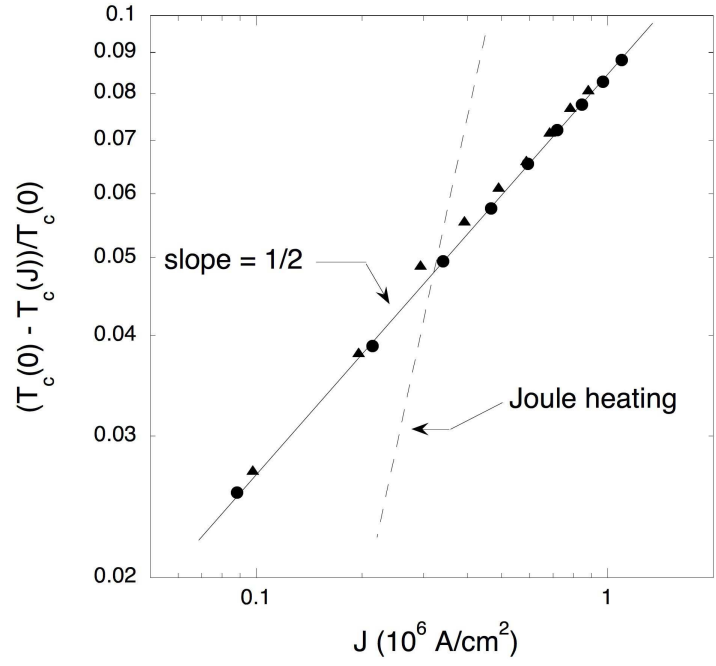}
\caption{\label{fig:epsart} Reduced transition temperature as a function of applied current density for the samples in Fig.\ 2.  The dashed
line has slope 2 and represents the effect of Joule heating.}
\newpage
\end{figure}

\begin{figure}
\includegraphics[width=5.in]{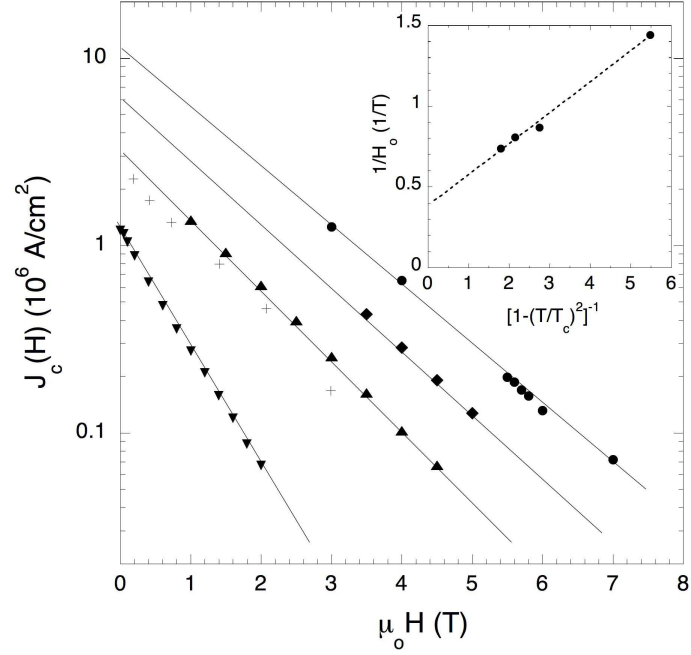}
\caption{\label{fig:epsart} Semi-log plot of the magnetic field dependence of the critical current density at several temperatures:
$\blacktriangledown$ 6.8 K, $\blacktriangle$ 6.0 K, $\blacklozenge$ 5.5 K, $\bullet$ 5.0 K, $+$ 5.0 K from Ref.  The field was applied
longitudinally to the fiber.  The solid lines represent exponential fits to the data from which a characteristic decay field $H_o$ is
extracted.  Inset: Inverse of characteristic field as a function of the inverse of the reduced temperature.  The dashed line is provided as a
guide to the eye.}
\newpage
\end{figure}



\end{document}